\documentstyle[prl,aps]{revtex}

\begin{document}

\title{Interaction of Conical Membrane Inclusions:
            Effect of Lateral Tension}
\author{T.~R.~Weikl $^{1),}$
         \thanks{Present address: Max-Planck-Institut f\"ur Kolloid- 
               und Grenzfl\"achenforschung, Kantstr.~55, 
               14513 Teltow, Germany},
           M.~M.~Kozlov $^{1),\,2)}$
             and  W.~Helfrich $^{1)}$ }  
\address{$^{1)}$ Freie Universit\"at Berlin,
          Fachbereich Physik,
          Arnimallee 14, 14195 Berlin, Germany \\
        $^{2)}$ Dept. Physiol. Pharmacol.,
               Sackler School of Medicine,
               Tel-Aviv University,
               Ramat-Aviv 69978, Israel}
\maketitle
\vspace*{.2cm}
\hspace*{2.7cm} {\small PACS Numbers: 87.22Bt, 87.10+e, 82.65Dp, 34.20-b} 
\begin{abstract}
Considering two rigid conical inclusions embedded in a membrane subject to
lateral tension, we study the membrane-mediated interaction between these 
inclusions that originates from the hat-shaped membrane deformations 
associated with the cones.
At non-vanishing lateral tensions, the interaction is found to depend on the  
orientation of the cones with respect to the membrane plane. 
The interaction of inclusions of
equal orientation is repulsive at all distances between them, 
while the inclusions of opposite orientation
repel each other at small separations, but attract each other at larger
ones. Both the repulsive and attractive forces become stronger with
increasing lateral tension.
This is different from what has been predicted on the basis of the same
static model for the case of vanishing lateral tension.
Without tension, the inclusions repel each other at all
distances independently of their relative orientation.
We conclude that lateral tension may induce
the aggregation of conical membrane inclusions.    
\end{abstract}

\section{Introduction}
Biological membranes consist of a fluid lipid bilayer with embedded 
amphiphilic macromolecules such as integral proteins \cite{theCell}. 
Integral proteins are expected to be much less flexible than the lipid 
matrix. 
In a general sense, any molecule embedded 
in the membrane and differing in shape or elastic properties from the 
surrounding lipid molecules can be viewed as an inclusion.
The phase behavior of inclusions in the plane
of the membrane is determined by interactions between them. 
If the interaction energy is sufficiently large to compete with translational
entropy, it can lead to lateral self-assembly of the inclusions. Attractive 
forces may result in a lateral aggregation of the inclusions, 
while repulsion can give rise to a regular array with 
maximal spacing.

Forces between membrane inclusions can be divided into two classes.
The first class consists of the well-known direct interactions, 
namely electrostatic (for charged inclusions) and van der Waals forces.
A second class comprises indirect interactions mediated by some kind of
membrane deformation \cite{Goulian1,Dan,Netz1,Goulian2,Park,Netz2,reviews}. 
These interactions are 
determined by the shapes of the inclusions and the elastic parameters
of the inclusions and the lipid bilayer. They can be static or dynamic,
in one case being due to equilibrium deformations and in the other to
shape fluctuations of the membrane. 

Both types of indirect interactions have been theoretically studied
for the case of zero lateral tension. The static interactions of inclusions
affecting the membrane thickness \cite{Dan} and of conical deformations 
affecting the membrane shape \cite{Goulian1} have been dealt with.
Dynamic interactions were treated for inclusions modifying the local 
bending moduli \cite{Goulian1,Netz1,Park,Netz2}, including the case of
rigid disks \cite{Goulian1,Goulian2,Park}.

In the following we consider the static interaction between conical
inclusions in the presence of lateral tension. Two sketches of a
truncated cone embedded in the membrane are given in Figs.~1 and 2.
The cone is assumed to be rigid and to impose a uniform slope on the
surrounding membrane which returns asymptotically to the flat state
at large distances. 

We find repulsive interaction at all values of the lateral tension 
if the two conical inclusions have equal orientation with respect to
the membrane plane. By increasing the 
lateral tension, the interaction is weakened at larger, but enhanced 
at smaller inclusion distances. In contrast, for opposite orientations of the 
inclusions in a membrane with non-vanishing lateral tension the sign of the
interaction
depends on the distance between the inclusions. At small separations the 
inclusions repel each other, while at large separations the interaction is 
attractive. With rising tension 
the attractive potential well deepens and moves towards smaller 
distances between the inclusions.

\section{Shape and energy of membranes with conical inclusions}
We consider a membrane with two embedded conical inclusions. 
The cross-sections
of the inclusions in the mid-plane of the membrane are
circles of radius $a$. The centers of the two circles 
are separated by the distance $R$ (see Fig.~3).

In the absence of inclusions the membrane is assumed to be flat and to
lie in the \mbox{$xy$-plane} of the Cartesian system of coordinates.
We describe the membrane equilibrium shape produced by the inclusions  
by a function $u(x,y)$ which determines the displacement of the membrane from
the $xy$-plane in $z$ direction (see Fig.~2).       

At the boundaries of the conical inclusions the displacement $u$
is assumed to fulfill the conditions (cf.~\cite{Goulian1}) 
\begin{equation}
u\bigg|_{r_i=a}=h_i+a\, \beta_i \cos\phi_i\;,\;\;\;\;\;\;
\frac{\partial u}{\partial r_i}\bigg|_{r_i=a}=\alpha_i+\beta_i \cos\phi_i\;,
\hspace{1cm}i=1,2    \label{1}
\end{equation}
where the subscript $i$ takes the values 1 or 2 for the first and the
second inclusion, respectively. By
$r_i$ and $\phi_i$ we denote the polar coordinates related to the center 
of projection $E_i$ of the respective inclusion on the $xy$-plane 
(see Fig.~3).  
The first equation in (1) describes the boundary of each inclusion 
as a circle of radius $a$
whose center is at height $h_i$ above the $xy$-plane
and which is tilted with respect to the $z$ axes by 
an angle $\beta_i$ in the $x$-direction. 
The second equation in (1) takes into account that due to the conical
shape of the inclusion the membrane is attached
to the circumference of the tilted circle with a constant angle $\alpha_i$.
It is assumed in (1) that the contact angle $\alpha_i$ and tilt
angle $\beta_i$ are small, $\alpha_i \ll 1, \beta_i \ll 1$,
so that we set $\tan\alpha_i = \alpha_i$, $\tan\beta_i = \beta_i$ and  
neglect contributions of the order of magnitude of $\beta_i^2$ 
determining the deviation of the inclusion projection from the circular shape.
At large distances from the inclusions, $r_i \gg R$, the membrane remains
flat, so that \mbox{$\mbox{\boldmath$\nabla$} u \to 0$} for $r_i \to \infty$.

The inclusions characterized by small $\alpha_i$ and $\beta_i$ can produce 
only a weak deformation of the initially flat membrane, which means that the 
gradient of the function $u(x,y)$ remains small, $|\nabla u| \ll 1$, everywhere
along the membrane.    
The membrane energy \cite{Helfrich} can then be written in the approximate 
form 
\begin{equation}
 G=\int\left(\frac{\kappa}{2}(\Delta u)^2 + \bar{\kappa} K 
        + \frac{\gamma}{2} (\mbox{\boldmath$\nabla$} u)^2\right)d^2r\;, 
      \label{2}
\end{equation}
where $\kappa$ denotes the bending rigidity, $\gamma$ the lateral tension,
$K$ the Gaussian curvature and $\bar{\kappa}$ the modulus of the Gaussian
curvature. In our approximation, the Laplacian
$\Delta u$ equals the sum of the principal curvatures  of the membrane 
$J$, while $\frac{1}{2}(\mbox{\boldmath$\nabla$} u)^2$ 
gives the increase of membrane area per unit projected area due to membrane
tilt $\mbox{\boldmath$\nabla$} u$.
The integration of (2) is performed 
over the projected area.
The membrane shape is determined by the Euler--Lagrange  
equation following from (\ref{2})
\begin{equation}
\Delta\Delta u = \frac{\gamma}{\kappa} \Delta u \;\;.\label{3}
\end{equation}

We derive the interaction energy of two conical inclusions in two
steps. First, we solve the shape equation (\ref{3}) accounting for the  
boundary conditions (\ref{1}) and the asymptotic boundary condition 
\mbox{$\mbox{\boldmath$\nabla$} u \to 0$} for $r_i \to \infty$. Second, 
inserting the obtained function $u(x,y)$ into (\ref{2}), we determine the 
membrane energy. Throughout this calculation we assume the 
inclusion distance $R$ to be large compared to the radius of inclusion $a$ 
and retain only the leading terms in $a/R$.

It is important to note that the interaction energy cannot depend on the
modulus of Gaussian curvature $\bar{\kappa}$. According to the theorem of 
Gauss-Bonnet an integral of the Gaussian curvature $K$ over a surface 
is equal to the 
negative sum of the line integrals of the geodetical curvature $k_g$
over the surface boundaries (apart from a constant that depends only 
on the genus of the surface). The value of the geodetical curvature at the 
inclusion boundaries is completely determined by the radius $a$ and contact
angles $\alpha_i$ ($|k_g| =1/a\cdot\cos\alpha_i$) and does not depend on the
distance $R$ between the inclusions. Therefore, the integral of the Gaussian 
curvature $K$ over the 
membrane must be independent of the distance $R$ and 
does not contribute to the interaction potential.  

For any given distance $R$ between the inclusions the energy has to be 
minimized with respect to the heights $h_i$ and tilt angles $\beta_i$.
This results in conditions of zero vertical force 
and zero torque acting on each
inclusion. The two conditions are expressed by the equations 
(see Appendix A)
\begin{eqnarray}
\int\limits_0^{2\pi}
        \left[a\frac{\partial}{\partial r_i}\left(
       \gamma u -\kappa \Delta u\right)\right]_{r_i=a} d\phi_i&=&0
              \nonumber  \\[.3cm]
\int\limits_0^{2\pi}\cos\phi_i
        \bigg[a^2\frac{\partial}{\partial r_i}\left(
       \gamma u -\kappa \Delta u\right)
        +\kappa a \Delta u\bigg]_{r_i=a} d\phi_i&=&0
       \hspace{.2cm},\label{4}
\end{eqnarray}
respectively.
The integration in (\ref{4}) is performed over the boundary of 
each inclusion.

\section{Interaction in absence of lateral tension}

We first consider the important limiting case of zero lateral tension,
$\gamma = 0$. The shape equation (\ref{3}) then reads 
\begin{equation}
\Delta\Delta u=0
\label{5}\;.
\end{equation}
To solve this equation satisfying the boundary conditions (\ref{1}) 
and the conditions of equilibrium (\ref{4}) we use the following 
ansatz. We consider the function $u(x,y)$ describing the shape of the membrane
in the form
\begin{equation}   
u=u_1(r_1,\phi_1) + u_2(r_2,\phi_2)\;, \label{6} 
\end{equation}
where $r_i, \phi_i$ denote polar coordinates with respect to
the center of inclusion $i$. The relationships between the 
polar coordinates 
related to the first and the second inclusion are (see Fig.~3)
\begin{eqnarray}
r_1&=&\sqrt{R^2+r_2^2-2Rr_2\cos{\phi_2}}\label{7}\\[.3cm]
\cos{\phi_1}&=&\frac{r_2\cos{\phi_2}-R}{\sqrt{R^2+r_2^2-2Rr_2\cos{\phi_2}}}
\label{8}
\end{eqnarray}
The functions $u_1$ and $u_2$ in (\ref{6}) are general solutions of the  
shape equation (\ref{5}) in polar coordinates. They are obtained 
from (\ref{a2}) derived in Appendix B and
have the form: 
\begin{eqnarray}
u_i(r_i,\phi_i)\!&=&\! \mbox{const.}+c_0^{(i)}\ln r_i + c_1^{(i)}r_i\cos\phi_i
            + c_2^{(i)}r_i\ln r_i \cos\phi_i+ c_3^{(i)}\frac{\cos\phi_i}{r_i}
             \nonumber\\
      &&  +\;c_4^{(i)}\cos 2\phi_i+ c_5^{(i)}\frac{\cos 2\phi_i}{r_i^2}+\ldots
         +c_{2n}^{(i)}\frac{\cos n\phi_i}{r_i^{n-2}}
         +c_{2n+1}^{(i)}\frac{\cos n\phi_i}{r_i^{n}}+\ldots\label{9}
\end{eqnarray}
Terms of (\ref{a2}) proportional to $\sin{n\phi}$ are omitted in 
(\ref{9}) because of the mirror symmetry
of the system with respect to the $xz$-plane (see Fig.~3), and terms 
exhibiting higher than logarithmical divergence for $r_i \to \infty$
are left out since they
violate the boundary condition of an asymptotically flat membrane 
\mbox{$\mbox{\boldmath$\nabla$} u \to 0$} for $r_i \to \infty$.
The only exceptions are $r_i\ln r_i\cos\phi_i$ and $r_i\cos\phi_i$.
From the boundary condition of  
an asymptotically flat membrane it can be concluded immediately
that the coefficient
$c_2^{(1)}$ must be equal to $-c_2^{(2)}$. 
The sum of the corresponding terms then diverges only logarithmically 
for $r_i \to \infty$.
The terms $r_i\cos\phi_i$ are proportional to the 
Cartesian coordinate $x$ and thus describe rotations 
of the membrane as a whole \cite{Bemerkung2}.
Any such rotations must be equal but opposite, i.e.~$c_1^{(1)}=-c_1^{(2)}$,
to satisfy the boundary conditions at infinity, so that we can as well drop 
these terms.
 
The coefficients $c_j^{(i)}$ in (\ref{9}) are determined from the 
boundary conditions (\ref{1})  and equilibrium conditions (\ref{4}).
Consider these conditions at the circumference of inclusion 2.
To apply them we have to express the membrane shape (\ref{6}) in the vicinity
of the inclusion.  The function $u_2$ is simply given by (\ref{9}) 
with $i=2$. 
To present the function $u_1$ in a convenient form we take
(\ref{9}) with $i=1$ and insert (\ref{7}) and (\ref{8}) into it. In the
vicinity of the second inclusion the value of $r_2$ is close to the inclusion
radius $r_2 \approx a$.
Using the assumption $a\ll R$ and, consequently, $r_2\ll R$  we perform a 
Taylor expansion about the center of the inclusion projection $E_2$
\begin{eqnarray}
u_1\bigg|_{r_2\ll R}&=&
       \mbox{const.}+c_0^{(1)}\left(\ln R-\frac{r_2}{R}\cos\phi_2
            -\frac{r_2^2}{2 R^2}\cos{2\phi_2}\right)
          - c_3^{(1)}\left(\frac{1}{R}+\frac{r_2}{R^2}\cos\phi_2\right)
                \nonumber\\
    && + c_2^{(1)}\left(-R\ln R+(1+\ln R)r_2\cos\phi_2
     -\frac{r_2^2}{2 R}+\frac{r_2^3}{12 R^2}(\cos 3\phi_2-3\cos\phi_2)\right)
          \nonumber\\
      && + c_4^{(1)}\left(1-\frac{r_2^2}{R^2}(1-\cos 2\phi_2)\right)
          + \frac{c_5^{(1)}}{R^2}
            -c_6^{(1)}\left(\frac{1}{R}+\frac{r_2}{R^2}\cos\phi_2\right)
            +\frac{c_8^{(1)}}{R^2} + O\left(\frac{r_2^3}{R^3}\right)
\end{eqnarray}  
The resulting expression for the membrane shape, $u = u_1+u_2$ is
\begin{equation}
 u\bigg|_{r_2\ll R} = f_0^{(2)}(r_2) +  f_1^{(2)}(r_2)\cos\phi_2 
            +  f_2^{(2)}(r_2)\cos 2\phi_2   
             +  f_3^{(2)}(r_2)\cos 3\phi_2 +\ldots\, \label{11} 
\end{equation}
where 
\begin{eqnarray}
f_0^{(2)} &=& \mbox{const.} + c_0^{(2)}\ln r_2  
          -r_2^2\left(\frac{c_2^{(1)}}{2 R} +\frac{c_4^{(1)}}{R^2}\right)
                 \nonumber\\
f_1^{(2)} &=& c_2^{(2)}r_2\ln r_2 + \frac{c_3^{(2)}}{r_2}
           +r_2 \left( -\frac{c_0^{(1)}}{R}
             +c_2^{(1)}\left(1+\ln R-\frac{r_2^2}{4 R^2}\right)
            -\frac{c_3^{(1)}+c_6^{(1)}}{R^2}\right)\nonumber\\
f_2^{(2)} &=& c_4^{(2)}+\frac{c_5^{(2)}}{r_2^2}+
            \left(-c_0^{(1)}+2c_4^{(1)}\right)
              \frac{r_2^2}{2R^2}\nonumber\\
f_3^{(2)} &=& \frac{c_6^{(2)}}{r_2}+\frac{c_7^{(2)}}{r_2^3}
              + \frac{c_2^{(1)}r_2^3}{12 R^2}
\end{eqnarray}
Inserting (\ref{11}) into the boundary conditions (\ref{1}) and 
equilibrium conditions (\ref{4}) at the
inclusion 2 we obtain a series of equations for the coefficients $c_j^{(i)}$. 

To account for the boundary and equilibrium conditions at the inclusion 1, we
perform the same procedure as described above to obtain identical equations 
in which the index 2 is replaced by 1 and 
{\it vice versa}.   

The equations obtained for the coefficients $c_j^{(i)}$  
can be solved order by 
order in the small parameter $a/R$. The solutions are
\begin{equation}
c_0^{(1)}=\alpha_1 a+O\left(\frac{1}{R^3}\right)\;,\;\;
 c_4^{(1)}=\frac{\alpha_2 a^3}{R^2}+O\left(\frac{1}{R^3}\right)
\;,\;\;c_5^{(1)}=-\frac{1}{2}\frac{\alpha_2 a^5}{R^2}
             +O\left(\frac{1}{R^3}\right) \label{12}
\end{equation}
and equivalent results for $c_j^{(2)}$, the remaining coefficients being
of third or higher order in $a/R$. 

We are now in a position to compute the energy of the membrane. 
Omitting the contribution of the integral of the Gaussian curvature, 
which is independent of the distance $R$ between the inclusions (see above) 
we obtain from (\ref{2})
\begin{equation} 
G(R) = \int \frac{\kappa}{2} J^2 \,d^2r  \label{13}
\end{equation}
where the curvature $J$ is given by  
\begin{equation}
 J = \Delta u = -4 c_4^{(1)} \frac{\cos 2\phi_1}{r_1^2}
                -4 c_4^{(2)} \frac{\cos 2\phi_2}{r_2^2} + \ldots 
\end{equation}  
In the first non-vanishing order in $a/R$, the energy of interaction of the
inclusions is  
\begin{equation}\fbox{\rule[-5.2mm]{0cm}{1.3cm}$\hspace*{.5cm}\displaystyle 
    G(R)=4\pi\kappa(\alpha_1^2+\alpha_2^2)\frac{a^4}{R^4}
              +O\left(\frac{1}{R^5}\right)\hspace*{.5cm}$}\label{14}
\end{equation} 
According to (\ref{14}), the energy is positive and decays monotonically at
all values of the contact angles $\alpha_1$, $\alpha_2$ and all distances
between the inclusions $R$. This means that in the case of zero lateral
tension the interaction between the rigid conical inclusions is always 
repulsive. The result (\ref{14}) is in agreement with an earlier calculation 
\cite{Goulian1} which in addition predicts a contribution proportional to
$\bar{\kappa}$, the modulus of Gaussian curvature. We think that there should
be no $\bar{\kappa}$ term (see end of Section II).

\section{Interaction in presence of lateral tension}

We now extend the methods of the previous section to analyze the interactions
of inclusions embedded in a membrane subject to non-vanishing 
lateral tension $\gamma$.  
The shape equation (\ref{3})  can be written as
\begin{equation} 
\Delta\Delta u=\xi^2\Delta u
\label{15}, 
\end{equation}
where $\xi = \sqrt{\gamma/\kappa}$ has the dimension of a reciprocal length.
To find a solution of the shape equation satisfying the boundary conditions 
(\ref{1})  and equilibrium 
conditions (\ref{4}) we use, as in the previous section, ansatz 
(\ref{6}) with the functions $u_i(r_i,\phi_i)$ being general 
solutions of the shape equation (\ref{15}) in polar coordinates. 
These functions are taken from (\ref{a3}) derived in Appendix B 
and have the form:
\begin{eqnarray}
u_i&=& \mbox{const.}+c_0^{(i)}K_0(\xi r_i)+c_1^{(i)}r_i\cos\phi_i
       + c_2^{(i)}K_1(\xi r_i) \cos\phi_i
           + c_3^{(i)}\frac{\cos\phi_i}{r_i}\nonumber\\
  &&+c_4^{(i)}K_2(\xi r_i)\cos 2\phi_i
         + c_5^{(i)}\frac{\cos 2\phi_i}{r_i^2}+\ldots
         +c_{2n}^{(i)}K_n(\xi r_i)\cos n\phi_i
         +c_{2n+1}^{(i)}\frac{\cos n\phi_i}{r_i^n}+\ldots\label{16}
\end{eqnarray}
The coefficients of all terms of (\ref{a3}) proportional to $\sin{n\phi}$ are
taken equal to zero because of the symmetry of the system. 
Also, the coefficients of terms violating the boundary condition of 
asymptotically vanishing gradient of the displacement,
\mbox{$\mbox{\boldmath$\nabla$} u \to 0$} for $r_i \to \infty$, must be zero. 
As we do not 
consider rotations of the system we set $c_1^{(i)}=0$. In addition, we 
omit in (\ref{16}) the logarithmic term of  (\ref{a3}), as the related change 
of the area of the membrane would result in an infinite energy of the lateral
tension $\gamma$.

Equation (\ref{16}) transforms into (\ref{9}) in the limit of vanishing 
lateral tension, 
$\gamma\to 0$ (i.e.~\mbox{$\xi \to 0$}). This can be shown by inserting into 
(\ref{16}) the approximative expressions
of the Bessel functions $K_n(x)$ for small arguments $x$ 
\begin{eqnarray}
K_0(x)\approx -\ln x\,,\hspace{.3cm}
K_1(x)\approx\frac{1}{x}\,,\hspace{.3cm}
K_n(x)\approx\frac{(n-1)!}{2}\left(\frac{2}{x}\right)^n
      -\frac{(n-2)!}{2}\left(\frac{2}{x}\right)^{n-2} \;\mbox{for}\; n\ge 2  
     \label{17}
\end{eqnarray}

The coefficients $c_j^{(i)}$ in (\ref{16}) are determined by the 
boundary conditions (\ref{1}) and equilibrium conditions (\ref{4}) 
in a way similar to that described in the
preceding section. For example, we present $u_1$ in the vicinity of 
the inclusion 2 by inserting (7) and (8) and obtain after an expansion
in the small parameter $r_2/R$ 
\begin{eqnarray}
u_1\bigg|_{r_2\ll R}&=&
       c_0^{(1)}\left( K_0(\xi R) + \xi K_1(\xi R) r_2\cos\phi_2
               +\frac{1}{4}\xi^2 r_2^2
                \left(K_0(\xi R)+K_2(\xi R)\cos 2\phi_2\right)\right)
                     \nonumber\\
       && + c_2^{(1)}\bigg(-K_1(\xi R)
     -\frac{1}{2}\xi\left(K_0(\xi R)+K_2(\xi R)\right)r_2\cos\phi_2
               \nonumber\\
 &&\hspace*{1.4cm} -\frac{1}{8}\xi^2\left(K_1(\xi R)+K_3(\xi R)\right) 
             r_2^2\cos 2\phi_2
    -\frac{1}{4}\xi^2 K_1(\xi R) r_2^2\bigg)\nonumber\\
    &&  + c_4^{(1)}\bigg(K_2(\xi R)
   +\frac{1}{2}\xi\left(K_1(\xi R)+K_3(\xi R)\right)r_2\cos\phi_2
        \nonumber\\
   &&\hspace*{1.4cm}+\frac{1}{8}\xi^2\left(K_0(\xi R)+K_4(\xi R)\right)
           r_2^2\cos 2\phi_2  +\frac{1}{4}\xi^2 K_2(\xi R)r_2^2 
               \bigg)\nonumber\\
  &&   - c_3^{(1)}\left(\frac{1}{R}+\frac{r_2}{R^2}\cos\phi_2\right)
          + \frac{c_5^{(1)}}{R^2} +\ldots
\label{19}
\end{eqnarray}   
We have to stress that in this case the expansions 
up to the second order in $r_2/R$ are sufficient 
only if $\xi a<1$. For $\xi a \gg 1$, which is equivalent to the 
condition of strong lateral tension, the series (\ref{19})  
converges too slowly to be approximated by the sum of just a few Taylor terms. 
This can be seen from the asymptotic expansion of the functions 
$K_n(x)$ which for large arguments are proportional to $\exp(-x)/\sqrt{x}$ 
irrespectively of $n$.

Inserting the sum (\ref{16}) with $i=2$ and (\ref{19}) 
into the boundary conditions (\ref{1})
and the equilibrium conditions (\ref{4}) at inclusion 2, we obtain  
for $a\ll R$ and  $\xi a<1$ a set of linear equations for
the coefficients $c_j^{(i)}$. 
Applying the same procedure to satisfy the boundary and 
equilibrium conditions at the
inclusion 1 leads to an analogous set of equations.  
Solving all these equations
for $a\ll R$ and $\xi a<1$ we obtain
\begin{eqnarray}
c_0^{(1)}=-\alpha_1 a+\ldots\,,\hspace{.5cm}
c_2^{(1)}=-\frac{1}{2}\alpha_2 a(\xi a)^2 K_1(\xi R)+\ldots\,,\hspace{.5cm}
        \\[.2cm] \nonumber
c_4^{(1)}=-\alpha_2 a (\xi a)^2 K_2(\xi R)+\ldots \hspace*{3cm}\label{20}
\end{eqnarray}
and corresponding results for $c_0^{(2)}$, $c_2^{(2)}$ and $c_4^{(2)}$. 
The coefficients $c_3^{(i)}$ and $c_5^{(i)}$ are given by
the relations 
\begin{eqnarray}
c_3^{(i)}=-\frac{1}{2}c_2^{(i)}\xi a^2 K_2(\xi a)\,,\hspace{.8cm}
c_5^{(i)}=-\frac{1}{4}c_4^{(i)}\xi a^3 K_3(\xi a)\,. \label{21}
\end{eqnarray}

The interaction energy $G(R)$ of the inclusions is obtained by integration 
of $\frac{1}{2}\kappa (\Delta u)^2$ over the $xy$-plane. Transforming the
area integrals into line integrals over the boundaries of the 
inclusions as shown
in Appendix C we find the following dominant terms for 
small $a/R$ and $\xi a<1$
\begin{equation} 
\fbox{\rule[-7mm]{0cm}{1.6cm}$\displaystyle\hspace*{.5cm}
G(R)=2\pi\kappa\alpha_1\alpha_2(\xi a)^2 K_0(\xi R)
      +\pi\kappa(\alpha_1^2+\alpha_2^2)(\xi a)^4 K_2^2(\xi R)
            +\ldots \hspace*{.5cm} $}    \label{22} 
\end{equation}
In the limit of vanishing tension, $\gamma\to 0$ (i.e.~$\xi\to 0$), 
this expression
for the interaction energy  
coincides with (\ref{14}), 
as can be seen by expanding the Bessel functions for small arguments
according to (\ref{17}).  

The interaction energy (\ref{22}) depends on the relative orientation
of the conical inclusions. 
If the cones are oriented in the same direction, their
contact angles $\alpha_1$, $\alpha_2$ have the same sign. 
In this case the energy of 
interaction is always positive and decreases with increasing distance $R$. 
The repulsive potential is illustrated in Fig.~4 for two identical and 
equally oriented inclusions 
at different lateral tensions. By increasing the tension the interaction is 
weakened at large,  but enhanced at small inclusion distances. 

If the inclusions are oppositely oriented, the contact angles $\alpha_1$ and 
$\alpha_2$ have different signs and the energy of interaction behaves 
non-monotonically,
as illustrated in Fig.~5. The energy  has a minimum at a finite separation 
$R^*$ of the 
inclusions. This means that the forces between the inclusions 
change from repulsive to
attractive depending on the distance $R$. 
For distances shorter than $R^*$ the inclusions repel each other, while for  
separations $ R > R^*$ the interaction is attractive. 
With rising lateral tension the separation of zero
force, $R^*$, moves towards smaller values and the associated 
potential well deepens (see Fig.~5).

\section{Conclusion}

To summarize, we derived an interaction energy between two conical inclusions 
embedded in a fluid membrane subject to lateral tension. For this purpose,
we calculated the equilibrium shape of an almost flat membrane and its
bending energy in the presence of inclusions as a function of their distance. 
In contrast to the case of vanishing tension, this interaction depends on the 
orientations of the inclusions with respect to the membrane plane. 
For oppositely oriented inclusions the interaction changes from
repulsive to attractive as the separation increases, while
equally oriented inclusions repel each other at all 
distances. This is very different from the case of vanishing lateral tension
where the interaction of conical inclusions is always repulsive, independently
of relative orientation.
 
We did not consider in this study the contribution of thermal 
undulations of the membrane
to the interaction between the inclusions \cite{Goulian1,Goulian2,Park}. 
In the case of non-vanishing lateral tension this may be partially 
justified by 
the fact that the undulations are diminished by the tension. Moreover, others
have found for the case of zero tension that the static part of the interaction
exceeds the dynamic one for $\kappa(\alpha_1^2+\alpha_2^2)>3\, k T$ 
\cite{Goulian1}
or $1.5\, k T$ \cite{Goulian2,Park} where $k$ is Boltzmann's constant and 
$T$ is temperature.
If the tension-induced forces dominate, they should lead to 
interesting phase behavior of embedded inclusions.
For example, the attractive interaction between pairs of 
oppositely oriented conical inclusions may favor  
the formation of clusters with a regular structure where inclusions with 
different signs of the contact angles alternate.
For an estimate of the attractive interaction one may use Fig.~5.
With $\kappa=1\cdot 10^{-19} J$ (typical of lipid bilayers),
$\alpha=0.5$ ($26.8^o$) and $\xi a =0.4$, the minimum of the interaction
energy $G(R)=V(R)\cdot \alpha_1^2 \kappa$ is roughly $-4 \cdot 10^{-21} J$  
($\approx k T$ at room temperature). Because of $\xi = \sqrt{\gamma/\kappa}$
the lateral tension needed to produce $\xi a = 0.4$ is given by
$\gamma = 0.16\, \kappa/a^2$. For $a=4\,nm$ and $\kappa=1\cdot 10^{-19} J$, one
finds $\gamma=1\,mN/m$ which is below the known tension of lipid bilayer 
rupture \cite{Evans}.

While our results are intuitively appealing and may be 
obtainable more directly,
we performed a complete perturbation calculation to make sure that no terms are
missed. The boundary and equilibrium conditions
for the membrane with conical inclusions are central to our calculations.
They resulted in a set of linear equations
for the coefficients of two superimposed expansions, 
one for either inclusion.    
This method is extendable to larger number of inclusions by using
similar sets of boundary conditions. In computer-aided calculation
the shape of a membrane could be determined with any desired precision and for 
any number of inclusions.

\begin{appendix}
\section{Derivation of the equilibrium conditions (4)}
To derive the equilibrium conditions (4) we study a variation
\begin{equation}
      v(r,\phi,\epsilon)=  u(r,\phi) + \epsilon\cdot\delta u(r,\phi) \label{V1}
\end{equation}
of the equilibrium membrane displacement $u(r,\phi)$ on a
circular ring S: $a\le r\le b,\; 0\le \phi\le 2\pi$ around 
a conical inclusion. To simplify the notation we leave out indices
of the polar coordinates $r,\phi$. The variation is restricted by the
boundary conditions (1) of the inclusion. So
\begin{eqnarray}
 \delta u\bigg|_{r=a} &=& \delta c + \delta\beta\, a\, \cos\phi\label{V2}\\
\frac{\partial\, \delta u}{\partial r}\bigg|_{r=a} &=& \delta\beta\,
            \cos\phi\label{V3}
\end{eqnarray}
where $\epsilon\cdot\delta c$ and  $\epsilon\cdot\delta\beta$ denote 
the changes of the hight of the inclusion center and the tilt angle,
respectively. At $r=b$ we set 
\begin{equation}
 \delta u\bigg|_{r=b} \,=\, \frac{\partial \delta u}{\partial r}\bigg|_{r=b} 
     \,=\,
                                                 0.\label{V4}
\end{equation}
Omitting Gaussian curvature the membrane energy (2) can be written as
\begin{eqnarray}
 G &=&\hspace{-.2cm} 
    \int_S\left(\frac{\kappa}{2}(\Delta v)^2 + 
      \frac{\gamma}{2}(\mbox{\boldmath$\nabla$}v)^2\right)
      \,d^2r\nonumber\\
  \hspace{-.2cm} &=& \hspace{-.2cm}
      \int\limits_0^{2\pi}\!\int\limits_a^b\left[\frac{\kappa}{2}
     \left(\frac{\partial^2 v}{\partial r^2}+\frac{1}{r}\frac{\partial v}
     {\partial r} + \frac{1}{r^2}\frac{\partial^2 v}{\partial \phi^2}
     \right)^2 + \frac{\gamma}{2}\left(\left(\frac{\partial v}
     {\partial r}\right)^2 + \frac{1}{r^2}\left(\frac{\partial v}
     {\partial \phi}\right)^2\right)\right]r\,dr\,d\phi\nonumber\\
  \hspace{-.2cm} &=&\hspace{-.2cm}
     \int\limits_0^{2\pi}\!\int\limits_a^b f(v,v_r,v_{\phi},
     v_{rr},v_{\phi\phi},r)\,dr\,d\phi\label{V5}
\end{eqnarray}
In equilibrium the energy $G$ is minimal. So
\begin{eqnarray}
\frac{dG}{d\epsilon}\bigg|_{\epsilon=0} = 
      \int\limits_0^{2\pi}\!\int\limits_a^b 
     \bigg[\frac{\partial f}{\partial v}\frac{dv}{d\epsilon} +
        \frac{\partial f}{\partial v_r}\frac{dv_r}{d\epsilon} +
        \frac{\partial f}{\partial v_\phi}\frac{dv_\phi}{d\epsilon} 
     +  \frac{\partial f}{\partial v_{rr}}\frac{dv_{rr}}{d\epsilon} +
     \frac{\partial f}{\partial v_{\phi\phi}}\frac{dv_{\phi\phi}}{d\epsilon}
       \bigg]_{\epsilon=0} dr\,d\phi 
        = 0 \label{V6}
\end{eqnarray}
By partial integrations we obtain
\begin{eqnarray}
\frac{dG}{d\epsilon} &=& \int\limits_0^{2\pi}\!\int\limits_a^b
  \left(\frac{\partial f}{\partial v}  
     - \frac{\partial}{\partial r}\frac{\partial f}{\partial v_r}
     - \frac{\partial}{\partial \phi}\frac{\partial f}{\partial v_\phi}
  + \frac{\partial^2}{\partial r^2}\frac{\partial f}{\partial v_{rr}}
   +\frac{\partial^2}{\partial\phi^2}\frac{\partial f}{\partial v_{\phi\phi}}
   \right)\frac{dv}{d\epsilon}\,dr\,d\phi\nonumber\\
&& +{}\int\limits_0^{2\pi}\left[
   \frac{\partial f}{\partial v_r}\frac{dv}{d\epsilon}
   -\left(\frac{\partial}{\partial r}
      \frac{\partial f}{\partial v_{rr}}\right)\frac{dv}{d\epsilon}
   +\frac{\partial f}{\partial v_{rr}}\frac{dv_r}{d\epsilon}
    \right]_a^b d\phi \label{V7}
\end{eqnarray}
and, inserting $f(v,v_r,v_{\phi},
     v_{rr},v_{\phi\phi},r)$ as defined in (\ref{V5}), are led to:
\begin{eqnarray}
\frac{dG}{d\epsilon} = \int\limits_0^{2\pi}\!\int\limits_a^b
   \left[\kappa\Delta\Delta v - \gamma \Delta v\right]
     \frac{dv}{d\epsilon}r\,dr\,d\phi 
     +\int\limits_0^{2\pi}\left[r\frac{\partial}{\partial r}\left(
       \gamma v -\kappa \Delta v\right)\frac{dv}{d\epsilon}
       +\kappa r \Delta v \frac{dv_r}{d\epsilon}\right]_a^b d\phi 
             \label{V8}
\end{eqnarray}
The equilibrium displacement $u(r,\phi)$ fulfills the shape equation (3)
of a tense membrane. So the integrand of the area integral in (\ref{V8})
is zero at $\epsilon=0$. Taking into account (\ref{V2}) and (\ref{V3}) 
we conclude 
\begin{eqnarray}
\delta G=\frac{dG}{d\epsilon}\bigg|_{\epsilon=0}
      =\int\limits_0^{2\pi}\bigg[r\frac{\partial}
  {\partial r}\left(\gamma u -\kappa \Delta u\right)
           (\delta c+\delta\beta \,a\,\cos\phi)\hspace{0cm}
         +\kappa r \Delta u\,\delta\beta\,\cos\phi) 
      \bigg]_{r=a} d\phi=0 \hspace{0cm}       \label{V9}
\end{eqnarray}
Since $\delta\beta$ and $\delta c$ are independent of each other we arrive
at the equations
\begin{eqnarray}
\delta G(\delta c)&=&\delta c\int\limits_0^{2\pi}
        \left[a\frac{\partial}{\partial r}\left(
       \gamma u -\kappa \Delta u\right)
          \right]_{r=a} d\phi
       =0 \hspace{.5cm}\label{V10}\\[.3cm]
\delta G(\delta \beta)&=&\delta \beta\int\limits_0^{2\pi}\cos\phi
        \bigg[a^2\frac{\partial}{\partial r}\left(
       \gamma u -\kappa \Delta u\right)
 +\kappa a \Delta u\bigg]_{r=a} d\phi
       =0\label{V11} 
\end{eqnarray}      
which state that the vertical force and the torque, respectively,  acting
on the inclusion must be zero in equilibrium.

\section{General solutions of the shape equations in polar coordinates}

In this appendix we derive the general solution of the shape equation 
\mbox{$\Delta\Delta u=\xi^2\Delta u$} 
in polar coordinates. We perform the calculation in two steps.
We first look for the solution  $J(r,\phi)$ of an intermediate equation 
$\Delta J=\xi^2 J$ and then solve the
equation $\Delta u=J(r,\phi)$. General solutions of the last equation are 
also general solutions of the shape equations. 

Below we consider separately the case of $\xi = 0$, 
corresponding to the vanishing
lateral tension, $\gamma = 0$, and the case of non-vanishing $\xi$.
\subsection{Vanishing lateral tension, $\xi = 0$}
In this case the shape equation has the form $\Delta\Delta u=0$.
A solution of the intermediate Laplace equation $\Delta J=0$ on a circular 
ring can be found by the method of separation of 
variables and reads \cite{Jackson,Pinsky}:
\begin{eqnarray}
J(r,\phi)=a_0 + b_0 \ln r + 
   \sum_{n=1}^\infty\left(a_n\cos n\phi +b_n\sin n\phi \right)r^{-n}
     +\sum_{n=1}^\infty\left(c_n\cos n\phi 
   +d_n\sin n\phi \right)r^n \label{a1}
\end{eqnarray}        
The general solution of the linear inhomogeneous 
equation  $\Delta u=J(r,\phi)$ is
the sum of a special solution and the general solution of 
the homogeneous equation
$\Delta u=0$, the latter having the form of (\ref{a1}). We obtain  
 \begin{eqnarray}
u(r,\phi)&=&A_0r^2+B_0r^2(\ln r-1)+(A_1\cos\phi +B_1\sin\phi )r\ln r
      \nonumber\\[.2cm]
      & & \hspace{-1cm}+{}\sum_{n=2}^\infty \left(A_n\cos n\phi +
                 B_n\sin n\phi\right)r^{-n+2}+
      \sum_{n=1}^\infty \left(C_n\cos n\phi +D_n\sin n\phi
                \right)r^{n+2}\nonumber\\[.1cm]
    \hspace*{1cm}  & &\hspace*{-1cm} +{}\bar{A}_0+\bar{B}_0\ln r 
+ \sum_{n=1}^\infty\left(\bar{A}_n
           \cos n\phi  +\bar{B}_n\sin n\phi \right)r^{-n}
           + \sum_{n=1}^\infty\left(\bar{C}_n\cos n\phi
        +\bar{D}_n\sin n\phi\right)r^n   \label{a2}
\end{eqnarray}
The terms with unbarred coefficients belong to the special solution, which
can be directly checked by its insertion 
into $\Delta u=J$. The $A_0$-term of (\ref{a2})   
corresponds to the $a_0$-term of (\ref{a1}) etc. 
The terms with barred coefficients
give the general solution of $\Delta u=0$ in analogy to (\ref{a1}). 

\subsection{Non-vanishing lateral tension, $\xi\ne$ 0}
By applying the method of separation of variables described in 
\cite{Jackson,Pinsky} 
also to the case of non-vanishing lateral tension we find 
the following general solution of the intermediate equation $\Delta J=\xi^2 J$
\begin{eqnarray*}
J(r,\phi)&=&  a_0 K_0(\xi r) + 
   \sum_{n=1}^\infty\left(a_n\cos n\phi +b_n\sin n\phi\right)
     K_n(\xi r)
 +\sum_{n=1}^\infty\left(c_n\cos n\phi
   +d_n\sin n\phi\right)I_n(\xi r)\;,
\end{eqnarray*}
where $I_n$ and $K_n$ denote modified Bessel functions.
A general solution of the equation $\Delta u=J(r,\phi)$ again consists of the 
sum of a special solution and the general solution (\ref{a1}) 
of the Laplace equation $\Delta u =0$. It can be written in the form
\begin{eqnarray}
u(r,\phi)&=& A_0 K_0(\xi r) + B_0 I_0(\xi R) +\bar{A}_0+\bar{B}_0\ln r
                  \nonumber\\
  &&\hspace*{-1cm} +\sum_{n=1}^\infty\left(A_n\cos n\phi+B_n\sin n\phi\right)
     K_n(\xi r)+\sum_{n=1}^\infty\left(C_n\cos n\phi
   +D_n\sin n\phi\right)I_n(\xi r)\nonumber\\    
 &&\hspace*{-1cm} + \sum_{n=1}^\infty\left(\bar{A}_n
           \cos n\phi +\bar{B}_n\sin n\phi\right)r^{-n}
   +\sum_{n=1}^\infty\left(\bar{C}_n\cos n\phi
        +\bar{D}_n\sin n\phi\right)r^n\;, \label{a3}
\end{eqnarray}
taking into account that $K_0(\xi r)$, $K_n(\xi r)\cos n\phi$, $I_0(\xi r)$,
 $I_n(\xi r)\cos n\phi$ and the corresponding terms containing $\sin{n\phi}$ 
are eigenfunctions
of the Laplace operator. 
\section{Reducing area integrals to line integrals in the calculation
            of the membrane energy}
In the calculation of the energy of the tense membrane  
\begin{equation}
G=\int \left[\frac{\kappa}{2}(\Delta u)^2+\frac{\gamma}{2}
     (\mbox{\boldmath$\nabla$} u)^2\right]d^2r
\end{equation}
we encounter, due to our ansatz (\ref{16}), integrals of the form
\begin{equation}
I(f,g)=\int\limits_{\mbox{\boldmath$R$}^2/E_1\cup E_2}\left[\frac{\kappa}{2}
     \Delta f\Delta g+
    \frac{\gamma}{2}\mbox{\boldmath$\nabla$}f\cdot
    \mbox{\boldmath$\nabla$}g\right]d^2 r
\end{equation}
where either $f$ obeys $\Delta f=\gamma/\kappa\cdot f$, which is true
for the terms of (\ref{16}) containing a Bessel function,
or $g$ is a solution of $\Delta g=0$, or both (see Appendix B).
$E_i$ denotes the projection of the inclusion $i$ into the 
$xy$-plane (see Fig.~3). Applying a theorem of Green we may write
\begin{equation}
\int\limits_{\mbox{\boldmath$R$}^2/E_1\cup E_2}
   \hspace*{-.5cm}   
(\mbox{\boldmath$\nabla$}f\cdot\mbox{\boldmath$\nabla$}g)\;d^2 r
=- \hspace*{-.5cm} \int\limits_{\mbox{\boldmath$R$}^2/E_1\cup E_2}  
 \hspace*{-.5cm}
    f\Delta g \;d^2 r - \int\limits_0^{2\pi} f\frac{\partial g}
         {\partial r_1}\Bigg|_{r_1=a}\hspace{-.3cm} a\, d\phi_1
  - \int\limits_0^{2\pi} f\frac{\partial g}
         {\partial r_2}\Bigg|_{r_2=a}\hspace{-.3cm} a\, d\phi_2
\end{equation}
if $r_i f(\partial g/\partial r_i)$ goes to zero for $r_i\to\infty$.
Since
\begin{equation}
\int\limits_{\mbox{\boldmath$R$}^2/E_1\cup E_2}\left[\frac{\kappa}{2}
 \Delta f\Delta g-\frac{\gamma}{2}f\Delta g 
  \right]d^2r\;=\;0\hspace{1cm}
\end{equation}
for  $\Delta f=\gamma/\kappa\cdot f$ or $\Delta g=0$, we find
\begin{equation}
I(f,g)=-\frac{\gamma}{2}\int\limits_0^{2\pi} f\frac{\partial g}
         {\partial r_1}\Bigg|_{r_1=a}\hspace{-.3cm} a\, d\phi_1
  -\frac{\gamma}{2}\int\limits_0^{2\pi} f\frac{\partial g}
         {\partial r_2}\Bigg|_{r_2=a}\hspace{-.3cm} a\, d\phi_2 \;.
\end{equation}

\end{appendix}


\vspace*{4cm}
{\large\bf  Figure Captions:}\\[1.2cm]
\underline{Figure 1}:\\  
 Conical inclusion in a bilayer membrane\\[1.2cm]
\underline{Figure 2}:\\ 
 Idealization of a conical inclusion as rigid disc 
 of height $h_i$ and tilt angle $\beta_i$ making a
 uniform contact angle $\alpha_i$ with the surrounding membrane.
 The cross section contains the axis of the cone.\\[1.2cm]
\underline{Figure 3}:\\ 
$xy$-plane with inclusion projections $E_1$ and $E_2$\\[1.2cm]
\underline{Figure 4}:\\ 
Dimensionless interaction potential 
       $V(R)=G(R)/(\alpha_1^2\kappa)$  of two
equally oriented inclusions \\($\alpha_1=\alpha_2$)
as a function of the dimensionless distance $R/a$
 for $\xi a=$ 0.05, 0.1, 0.2 \\[1.2cm]
\underline{Figure 5}:\\ 
Dimensionless interaction potential $V(R)=G(R)/(\alpha_1^2\kappa)$
of two oppositely oriented inclusions ($\alpha_1=-\alpha_2$)
as a function of the dimensionless distance $R/a$. 
The potential well deepens with increasing 
$\xi a=$ 0.1, 0.2, 0.3, 0.4, 0.5. Note that
$R/a = 2$ means two discs in contact. Near this value the results
can only be regarded as estimates.


\begin{thebibliography}{99}
\bibitem{Bemerkung2} The shape $u$ of a tensionless membrane with a 
     single inclusion for example is just $u=\alpha a\ln r$. 
     Enforcing a tilt of the inclusion by an angle $\beta$ 
     results in the shape $u=\alpha a\ln r + \beta r\cos\phi$, 
     i.e.~a tilt of the whole membrane by $\beta$.
\end{thebibliography}
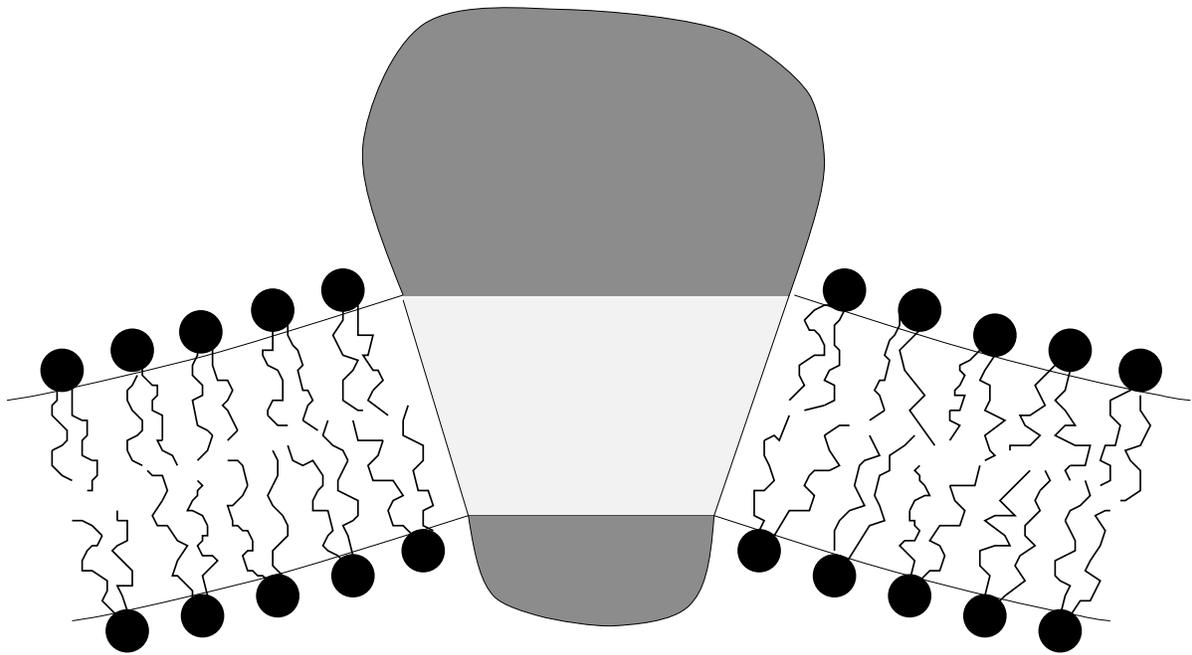
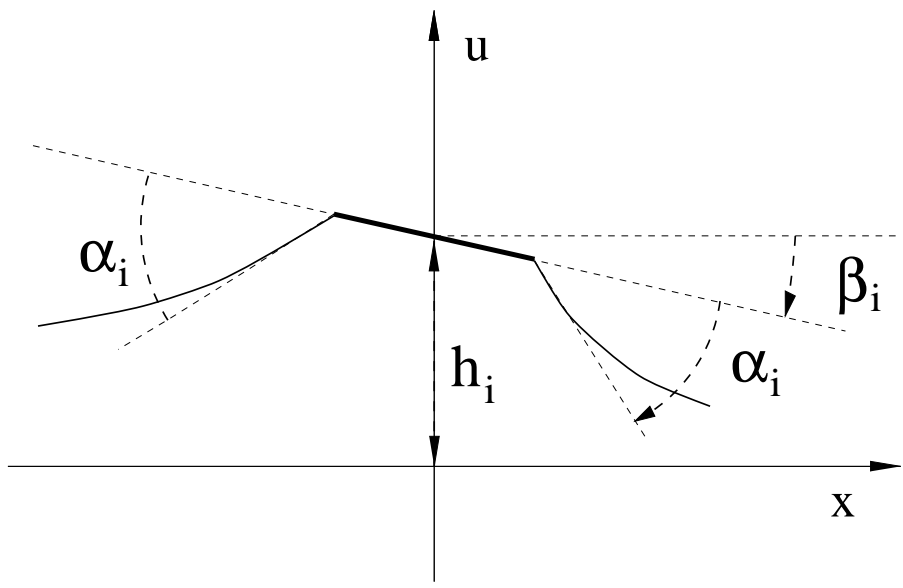
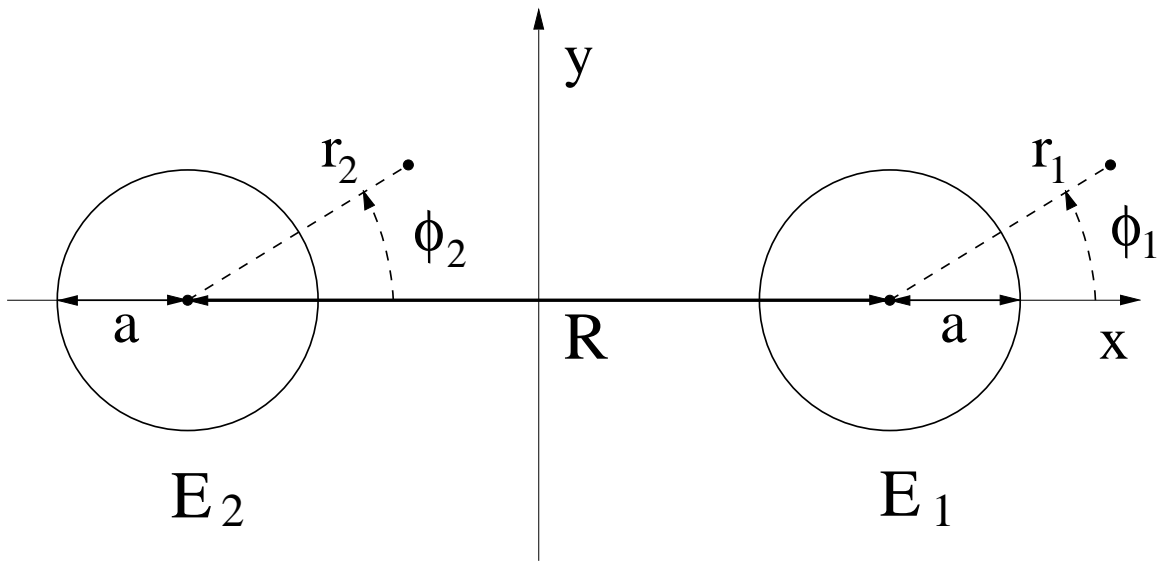
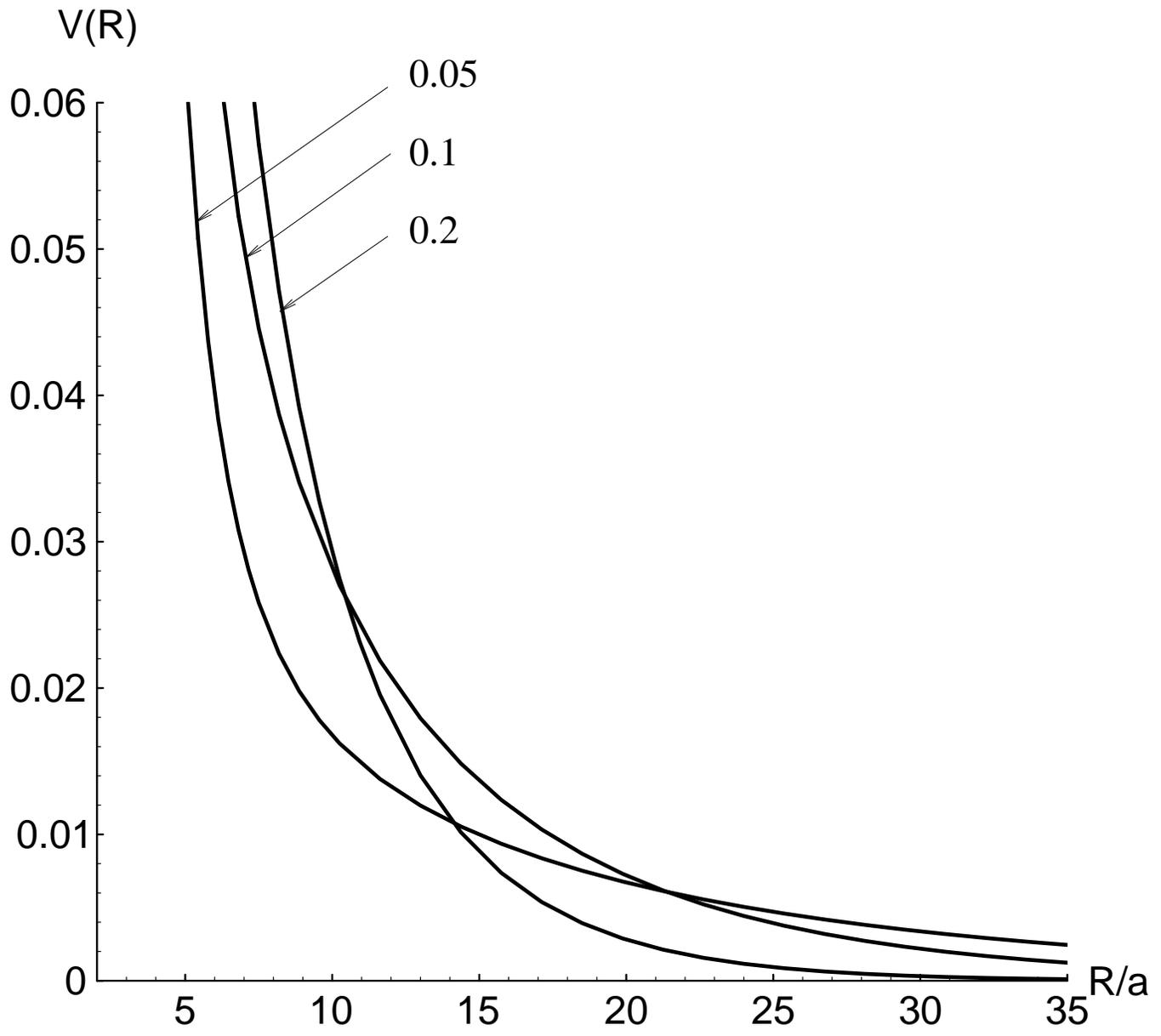
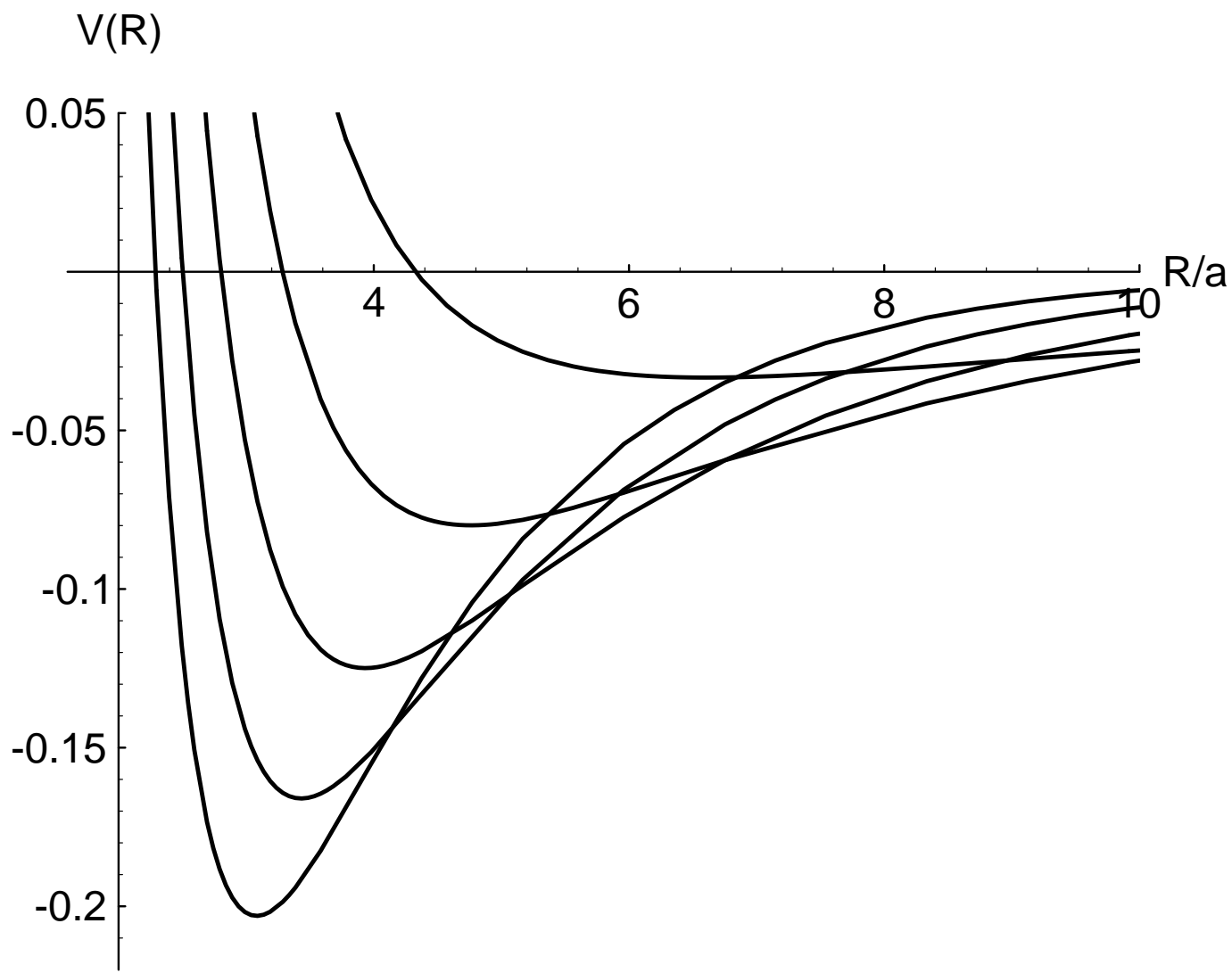
\end{document}